%% file: sample-sigconf.tex
\documentclass[sigconf]{acmart}
\usepackage{comment}
\usepackage{tikz}
\usepackage{hyperref}
\usepackage{cleveref}
\usepackage{subcaption}
\usepackage{booktabs}
\usepackage{tabularx}
\usepackage[]{todonotes} 
\AtBeginDocument{%
  }

\acmConference[]{Conference’23}{Date}{Place}

\begin{document}

\title{Mitigating Intersection Attacks in Anonymous Microblogging}


\author{Sarah Abdelwahab Gaballah}
\affiliation{%
  \institution{Telecooperation Lab (TK), Technical University of Darmstadt}
  \streetaddress{Hochschulstr. 10}
  \city{Darmstadt}
  \country{Germany}}
\email{gaballah@tk.tu-darmstadt.de}

\author{Thanh Hoang Long Nguyen}
\affiliation{%
  \institution{Telecooperation Lab (TK), Technical University of Darmstadt}
  \streetaddress{Hochschulstr. 10}
  \city{Darmstadt}
  \country{Germany}}
\email{long.nguyen@stud.tu-darmstadt.de}

\author{Lamya Abdullah}
\affiliation{%
  \institution{Telecooperation Lab (TK), Technical University of Darmstadt}
  \streetaddress{Hochschulstr. 10}
  \city{Darmstadt}
  \country{Germany}}
\email{abdullah@tk.tu-darmstadt.de}

\author{Ephraim Zimmer}
\affiliation{%
  \institution{Telecooperation Lab (TK), Technical University of Darmstadt}
  \streetaddress{Hochschulstr. 10}
  \city{Darmstadt}
  \country{Germany}}
\email{zimmer@privacy-trust.tu-darmstadt.de}

\author{Max M\"uhlh\"auser}
\affiliation{%
  \institution{Telecooperation Lab (TK), Technical University of Darmstadt}
  \streetaddress{Hochschulstr. 10}
  \city{Darmstadt}
  \country{Germany}}
\email{max@tk.tu-darmstadt.de}

\renewcommand{\shortauthors}{S. A. Gaballah}

\begin{abstract}
Anonymous microblogging systems are known to be vulnerable to intersection attacks due to network churn. An adversary that monitors all communications can leverage the churn to learn who is publishing what with increasing confidence over time. In this paper, we propose a protocol for mitigating intersection attacks in anonymous microblogging systems by grouping users into anonymity sets based on similarities in their publishing behavior. The protocol provides a configurable communication schedule for users in each set to manage the inevitable trade-off between latency and bandwidth overhead. In our evaluation, we use real-world datasets from two popular microblogging platforms, Twitter and Reddit, to simulate user publishing behavior. The results demonstrate that the protocol can protect users against intersection attacks at low bandwidth overhead when the users adhere to communication schedules. In addition, the protocol can sustain a slow degradation in the size of the anonymity set over time under various churn rates.
\end{abstract}




\maketitle

\input{sections/intro}

\input{sections/models}

\input{sections/solution}

\input{sections/eval}

\input{sections/diss}

\input{sections/relatedwork}

\input{sections/conclusion}

\bibliographystyle{ACM-Reference-Format}
\bibliography{references}

\end{document}

%% file: sections/intro.tex
\section{Introduction} 
Microblogging is a popular form of online social networking that enables the rapid dissemination of information and news. Platforms that support microblogging, such as Facebook and Twitter, have played a substantial role during sociopolitical protests and crisis situations, such as the 2022 Iran protests~\cite{intro4} or the 2023 Turkey-Syria earthquake~\cite{intro5}. However, freely expressing one's views on these platforms may have serious ramifications. An activist who is caught posting about a regime-critical topic, for example, may face serious legal consequences \cite{intro3}. Additionally, by observing which topics a user is publishing on, service providers can deduce sensitive information such as health issues, financial status, or sexual preferences. Creating fake accounts on these platforms is a popular strategy to hide real identities. However, this does not solve the problem because communication metadata, such as the user's IP address, can be used by the platforms to associate the fake account with the user's location or identity. 

Over the last years, many anonymous communication systems have been proposed to protect communication metadata, with some of these systems mainly designed for microblogging~\cite{atom,RefRiffle,riposte,RefBlinder,sarah}. An anonymous communication system can initially ensure that a user cannot be identified among a group of other users, known as an anonymity set \cite{pfitzmann}. However, the anonymity sets change over time due to network churn. This change in the anonymity sets makes users susceptible to traffic analysis attacks. 
 
Intersection attacks are one of the strongest traffic analysis attacks~\cite{openlimits,2ssda,berthold03,statistical,portela2,wolinsky13,signalattack}. These attacks could be applied against almost any existing practical anonymous communication system~\cite{hayes16}. In these attacks, an adversary who monitors the communication can intersect the anonymity sets over time to single out a certain user~\cite{berthold03}. An example of these attacks is when a corrupt mayor discovers that someone in the city has created a Facebook account with a fictitious name and exposes information about the mayor's corruption or illegal/immoral act. To determine who owns this account, the mayor forces the internet service provider of the city to provide him with a list of the names of people who are using Facebook (or connecting to the Internet) whenever a new post is published on the targeted account. Each list may contain many users; however, when the mayor intersects these lists, the size of the resulting set decreases over time until it contains only one user, the account's owner. These attacks are applicable even if the account's owner connects to Facebook using an anonymous communication system. 

Intersection attacks are very powerful, particularly when launched for a long time. Their mode of operation typically falls into the category of passive attacks, which means users will not become aware of the fact that an attack is taking place. They can either be performed deterministically, meaning that, in case the attack is successful, the adversary is able to link a user to her fake account with absolute certainty. Or they can be performed probabilistically---this variant is called a statistical disclosure attack---which aims at estimating the likelihood that a target user was the owner of a specific account among a group of users~\cite{hayes16}.
 
Many solutions have been introduced in the literature to mitigate the intersection attacks. These solutions include sending dummy/cover messages (also known as cover traffic) to hide the real communication~\cite{anonpubsub, RefMTor} or delaying the users' messages on the anonymous communication system side for a random amount of time~\cite{atom}. Supporting a wide userbase was also recommended in~\cite{riposte} as a strategy to increase anonymity sets and thereby hinder intersection attacks. Nevertheless, all these solutions have been shown to be ineffective~\cite{berthold03, grube, gaballah2022effectiveness}.  Anonymous communication systems, such as~\cite{RefBlinder,sarah, RefRiffle}, consider constant user participation, i.e., the requirement for users to always be online and send messages to the system, as the only way to effectively protect against intersection attacks. However, this requirement is not realistic or practical. A~framework for vulnerability monitoring and active mitigation of anonymity loss under intersection attacks was proposed in~\cite{wolinsky13}. Nevertheless, this framework incurs considerable bandwidth and latency costs due to the inefficient method it uses to build anonymity sets and the random assignment of users to fixed sets of the same size.

In this paper, we propose a protocol for protecting users who publish messages on an anonymous microblogging system from being de-anonymized (i.e., linked to their published content) by intersection attacks. For the sake of efficiency, the protocol works by forming anonymity sets based on the similarity of the users' publishing behavior. It creates a communication schedule for users in each anonymity set to control message transmissions in such a way that users within a set behave indistinguishably from the point of view of an adversary. The schedules can be adjusted to optimize the trade-off between bandwidth overhead and latency based on the users' performance needs. Our protocol focuses only on protecting users when they are publishers on an anonymous microblogging system, so it is out of our scope to protect them when they are subscribers.\footnote{In this case, users can be protected by broadcasting published messages, i.e., the system sends every published message to all users, as seen in~\cite{RefBlinder,RefDissent,riposte}. However, broadcasting imposes a significant communication overhead on users, making it an inefficient solution. As a result, more research in this area is still clearly required.}

The paper's main contributions are: (1) a protocol that prevents intersection attacks by grouping users into sets according to how they publish and establishing communication schedules that enforce indistinguishability across users in the same set; (2) an analysis of realistic user behavior with the help of real-world datasets from Twitter and Reddit; and (3) an evaluation of the protocol, in which we study the impact of the schedule design on bandwidth and latency, as well as the impact of the churn rate on the size of the sets provided by our protocol.

This paper is organized as follows: Section~\ref{sec:models} introduces the system and threat models. Section~\ref{sec:solution} then presents our protocol and its five phases. Following that, in Section~\ref{sec:eval}, we discuss the evaluation results of our experiments. Section~\ref{sec:diss} presents a discussion on some additional settings in our protocol. In Section~\ref{sec:rw}, we provide a review of the related work. Finally, Section~\ref{sec:confw} concludes the paper and presents future work.

%% file: sections/models.tex
\section{Models}
\label{sec:models}
This section describes the system model and the design assumptions of our mitigation protocol. It also discusses the adversary's capabilities and goals.

\subsection{System Model}
\label{sec:sm}
Our mitigation protocol is assumed to be employed by an anonymous microblogging system. We do not restrict the system to any particular anonymity technique. Instead, we consider a broadly applicable decentralized system based on an anytrust model, i.e., a system run by many servers (e.g., many mix nodes) where at least one of them is trustworthy \cite{RefRiffle}. Even if some of the system's servers are malicious, the system is assumed to be honest in its execution of the protocol.

The system allows users to publish posts under pseudonyms on a shared board (e.g., a public bulletin board). However, the users are able to publish their posts only when the protocol permits. The protocol is carried out for every set of new users $U$ (we call $U$ a "batch"), where the size of $U$ should be above a pre-defined and system specific value. Each user $u_i \in U=\{u_1,u_2,...u_n\}$ has only one pseudonym $p_j \in P=\{p_1,p_2,...p_n\}$ and vice versa. The system is responsible for ensuring \textit{unlinkability} between $u_i$ and $p_j$. 

The communication in the system is assumed to proceed in time intervals $T=\{T_1,T_2,...T_v\}$. Each time interval $T_e \in T$ consists of a~set of time slots $\{t_1,t_2,...,t_w\}$, where $|T_e|=w$. The users send their messages during the slots. Users can send at any time during the slot period, but the content of the messages will be published publicly by the system only at the end of the slot. 

All the exchanged messages between the users and the system should be encrypted, and padded to the same length. To prevent an adversary from probabilistically profiling a user based on rates of sending, every user that wants to send in a time slot $t_{l} \in T_e$ must send $m$ messages. To reach the required number of messages $m$, the user can send cover messages if the number of her actual messages is less than $m$. If a user has more than $m$ real messages in $t_{l}$, the extra messages should be delayed in a \textit{message queue} on the user's side where they can be sent later during the next slot(s).

When users send cover messages, these messages will not be part of the published content, as they are just used to feign identical communication behavior among the users. Since the content of users' real messages will be published publicly and can be read by anyone, users must not include any personally identifiable information, such as real names or addresses, in the content.


\subsection{Threat Model}
\label{sec:tm}
We assume a global passive adversary $\mathcal{A}$ who observes the whole communication. $\mathcal{A}$ can only see the message’s metadata but not the content. It does not alter, delay, or drop packets sent by users. Also, we do not consider the ability of $\mathcal{A}$ to launch Sybil attacks. Additionally, we assume that $\mathcal{A}$ cannot corrupt the functionality of the system or deny its availability. Moreover, it cannot control the whole system, thus it cannot break the unlinkability property that is provided by the system to link a user to her pseudonym. To de-anonymize users, $\mathcal{A}$ utilizes an intersection attack. It launches the attack by observing and analyzing the publishing behavior of the users. When there is no change in the behavior of users belonging to the same set, $\mathcal{A}$ fails to de-anonymize users, and hence intersection attacks are rendered ineffective. Any de-anonymization attack based on analyzing the published content---e.g., analyzing the writing styles---is out of our scope.

%% file: sections/solution.tex
\section{Protocol Architecture}
\label{sec:solution}
\begin{figure*}[t!]
   \centerline{\includegraphics[width=0.92\textwidth, height=12mm]{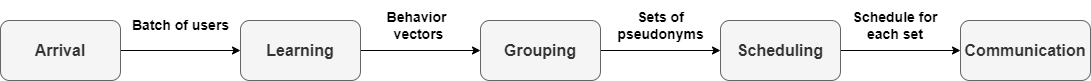}}
\caption{A process diagram of the protocol's phases }
\label{fig:pd}
\end{figure*}

In this section, we describe our mitigation protocol in detail. The protocol is divided into five phases: the \emph{Arrival Phase} (Section~\ref{sec:p1}), the \emph{Learning Phase} (Section~\ref{sec:p2}), the \emph{Grouping Phase} (Section~\ref{sec:p3}), the \emph{Scheduling Phase} (Section~\ref{sec:p4}), and the \emph{Communication Phase} (Section~\ref{sec:p5}). The protocol is carried out in batches, with each batch of new users going through its own set of phases. Figure~\ref{fig:pd} illustrates  a general process diagram of the protocol’s phases.

\subsection{Arrival Phase}
\label{sec:p1}
During this phase, the protocol waits until it collects a batch of new users $U$ who want to join the microblogging system. When the size of the batch $U$ reaches the pre-defined batch threshold, the users in the batch are notified to begin the \emph{Learning Phase}, during which they are able to communicate and publish their messages. If a user $u_i \in U$ has a message to publish, but the size of $U$ is still less than the threshold, the message should be delayed in the user's message queue and sent only to the system when $u_i$ enters the \emph{Learning Phase}. In Section~\ref{sec:diss}, we go into further depth about the delay that is introduced during the \emph{Arrival Phase}.

After a complete batch enters the \emph{Learning Phase}, the protocol can start a new \emph{Arrival Phase} to accumulate a new batch.

\subsection{Learning Phase}
\label{sec:p2}
In this phase, the protocol has a batch $U$, and needs to learn the publishing behavior of the users in $U$ during a time interval $T_{learning}$ (where $T_{learning}=T_1$), i.e., in which time slots the users send real messages to be published. Since the system that executes the protocol is not totally trusted (we assume an anytrust model, cf. Section~\ref{sec:sm}), the protocol is not able to learn the publishing behavior of the users directly. Instead, it learns the publishing behavior on the pseudonyms in $P$, i.e., in which time slots the pseudonyms have messages. This learned behavior is used to reflect the behavior of the users, who are the owners of the pseudonyms.

We refer to the publishing behavior on a pseudonym $p_j \in P$ as a binary vector $B_{p_j}\in \{0,1\}^w$. To learn the publishing behavior on pseudonyms, the protocol requires every $u_i \in U$ to send $m$ messages (real or cover) in every $t_l \in T_{learning}$. Each real message sent by the user should specify the actual slot when the message was created on the user's side. The message's creation slot may differ from the message's sending slot. That happens because the messages might be delayed on the user's side until the \emph{Learning Phase} starts or because the user has more messages than what is allowed to be transmitted in a slot.

The protocol computes $B_{p_j}$ in every $t_l$ as follows:
\begin{equation}
B_{p_j,t_l} =
  \begin{cases}
    1      & \quad \text{if } p_j \text{ has real messages created at } t_l \\
    0  & \quad \text{if } p_j \text{ has no real messages created at } t_l
  \end{cases}
  \label{eq:1}
\end{equation}
By the end of the \emph{Learning Phase}, the protocol has a binary vector for each pseudonym $p_j \in P$ which demonstrates the publishing behavior on this pseudonym during $T_{learning}$. In this vector, when a slot has a value of 1, it means the $p_j$'s owner created real message(s) during this slot, while a value of 0 indicates the opposite.

\subsection{Grouping Phase}
\label{sec:p3}
The goal, in this phase, is to group the pseudonyms in the set $P$ based on the similarity in their publishing behavior. To achieve this goal, a $k$-mode clustering algorithm is employed \cite{kmode}. We consider the $k$-mode algorithm due to its capability of grouping data points efficiently in such a way that minimizes the total mismatches between the corresponding attribute values of the two data points.
The~protocol performs clustering by carrying out the following steps.
\begin{itemize}
\item Randomly select a set of $k$ pseudonyms from $P$.~\footnote{There are some methods to determine the best value of $k$ such as the elbow method~\cite{elbow}.}  The publishing behavior vectors of the chosen pseudonyms are the initial cluster heads, with one vector assigned to each head. The set of clusters is defined as $S={S_1,S_2,...,S_k}$, and the head of a cluster $S_x \in S$ is denoted as $S_x^{head} \in \{0,1\}^w$.

\item Calculate the distance between every head $S_x^{head}$ and the publishing behavior vector of every pseudonym $p_j \in P$. The distance between the two vectors, $B_{p_j}$ and $S_x^{head}$, is computed as the total mismatches in each slot's value in the two vectors. The smaller the number of mismatches is, the less the distance is (i.e., the more similar the two vectors are). This distance measure is known as the Simple Matching Coefficient (SMC) \cite{kmode}. Formally, the distance is calculated as follows:
\begin{equation}
dist(B_{p_j}, S_x^{head}) = \sum_{l=1}^{w} \delta(B_{p_j,t_l}, S_{x,t_l}^{head})
\end{equation}
where
\begin{equation}
\delta(B_{p_j,t_l}, S_{x,t_l}^{head}) =
  \begin{cases}
    0      & \quad \text{if } B_{p_j,t_l}=S_{x,t_l}^{head}\\
    1  & \quad \text{if } B_{p_j,t_l} \neq S_{x,t_l}^{head}
  \end{cases}
\end{equation}

\item  Assign each pseudonym $p_j$ to a cluster $S_x$ whose head $S_x^{head}$ is the most similar to the behavior vector $B_{p_j}$.

\item Update $S_x^{head}$ for each cluster $S_x \in S$.  The new instance of $S_x^{head}$ is calculated as the mode of the behavior vectors of all pseudonyms in $S_x$. Thus, $S_{x,t_l}^{head} = 1$ if $t_l$ in most of the vectors has a value of 1; otherwise, $S_{x,t_l}^{head}$ is assigned a value of~0.  

\item Repeat steps 2-4 until there are no more changes in the clusters, i.e., no updates in the cluster heads and/or the members of the clusters. 

\end{itemize}

The output of the previous steps is a set of clusters $S$, where each cluster $S_x \in S$ consists of a set of pseudonyms that are similar in terms of publishing behavior vectors.

\subsection{Scheduling Phase}
\label{sec:p4}
To prevent $\mathcal{A}$ from de-anonymizing users using intersection attacks, all owners of pseudonyms belonging to a set $S_x$ must communicate in an indistinguishable manner. To achieve that, the protocol creates a schedule $H_x \in \{0,1\}^w$, that specifies the time slots on which the owners of the pseudonyms in $S_x$ should send their messages during the time intervals of the \emph{Communication Phase}. In our protocol, we construct the communication schedules for the users to communicate in the future based on their communication history. Initially, in this phase, the schedule $H_x$ is computed based on the publishing behavior vectors learned during the \emph{Learning Phase} (Section~\ref{sec:p2}). Later, in the \emph{Communication Phase} (Section~\ref{sec:p5}), we discuss how the schedules can be updated.

\begin{figure}[htbp]
\centerline{\includegraphics[width=0.64\columnwidth, height=54mm]{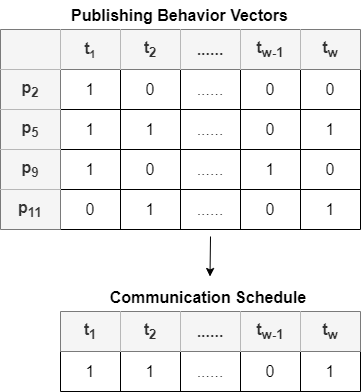}}
\caption{Example for the scheduling where the activity threshold is 50\%.}
\label{fig:example}
\end{figure}

This phase consists of two steps:
\begin{enumerate}
\item  \textit{Creation}: $H_x$ consists of $w$ time slots, where the value of each slot $t_{l} \in H_x$ is computed as:
\begin{equation}
H_{x,t_l} =
  \begin{cases}
    1      & \quad \text{if } \sum_{j=1}^{|S_x|} B_{p_j,t_l} \geqslant q \\
    \\
    0  & \quad \text{if } \sum_{j=1}^{|S_x|} B_{p_j,t_l} < q
  \end{cases}
   \label{eq:5}
\end{equation}
where $q$ refers to the activity threshold in each $t_{l}$.

Each time slot in the schedule $H_{x,t_l}$ is assigned a value of either 1 or 0 based on the total number of  publishing behavior vectors having $t_l$ with a value of 1 (i.e., activity).
During the \emph{Communication Phase}, each user who owns a pseudonym $p_j \in S_x$ must send a message during a slot $t_l$ when the value of $H_{x,t_l} =1$ (we refer to $t_l$ in this case as an \textit{active slot}). While $H_{x,t_l} = 0$ means that the user must not send any message during that slot, because $t_l$ in this case is \textit{inactive slot}. An example of how to create a schedule is shown in Figure~\ref{fig:example}.

\item \textit{Broadcasting}: After the schedules are created, all users receive information about the clustered pseudonyms and their schedules (obviously, pseudonyms that belong to the same cluster have the same schedule).
Each user identifies the corresponding cluster of her pseudonym and the related schedule. Since the protocol does not recognize the pseudonyms of the users, it cannot directly send each user her designated schedule.\footnote{The amount of overhead introduced by this broadcasting step depends on the batch size. The overhead can be avoided if the system allows anonymous retrieval, e.g., by using a private information retrieval technique \cite{RefRiffle, sarah}. Each user can then retrieve only the information pertinent to her schedule without the system being able to tell what the retrieved information is.}
\end{enumerate}

The bandwidth overhead and latency introduced by applying a schedule $H_x$ are highly influenced by the value of $q$. When it is low, many slots in $H_x$ may be assigned a value of 1. Consequently, the users will be required to send messages in many time slots during the \emph{Communication Phase}. Thus, the low value of $q$ may result in a high bandwidth overhead for the users during the \emph{Communication Phase}. When the $q$ value is high, i.e. the opposite case, it may not be easy to reach the threshold. Therefore, $H_x$ could contain only a low number of \textit{active slots}, probably resulting in a high latency for many users of the corresponding cluster during the \emph{Communication Phase}, especially for users with high publishing rates. In the evaluation section, we discuss the impact of $q$ on the bandwidth overhead and latency in greater detail.

\subsection{Communication Phase}
\label{sec:p5}
By the end of the \emph{Scheduling Phase}, every user $u_i \in U$ is assigned a schedule $H_x$, where $u_i$'s pseudonym $p_j \in S_x$ . The user $u_i$ must strictly follow the schedule, which means that $u_i$ should send messages when $H_{x,t_l} = 1$ and refrain from sending any message when $H_{x,t_l} = 0$. The number of messages that $u_i$ must send when $H_{x,t_l} = 1$ is $m$. If $u_i$ is scheduled to send in a specific time slot but does not have real messages to send, she should send cover messages. When $u_i$ creates real messages during $t_l$, and according to the schedule, she must not send in that time slot, the created messages shall be pushed to a message queue that resides on the user's side.  \\

When the users strictly follow the schedules that are assigned to them, their anonymity sets will be \textit{indistinguishability sets}.\\

\textbf{Definition.} \textit{$S'$ is an indistinguishability set if all users in this set have the same behavior when they send messages to the system. The probability for $\mathcal{A}$ guessing the pseudonym $p_j \in S_x$ of a user $u_i \in S'$ is $1/|S'|$.}\\

Intuitively, this means that $\mathcal{A}$ can de-anonymize $u_i$  (link a pseudonym to $u_i$) by only making random guesses. If one user in $S'$ slightly deviates from the behavior of the other users in the set, the protocol will not be able to guarantee indistinguishability for this user. The larger the size of the indistinguishability set $|S'|$ is, the more protected the users are. To guarantee a minimum level of indistinguishability, the protocol must ensure that the size of every indistinguishability set $|S'|$ is larger than a certain number $z$. \\

\paragraph*{\textbf{Churn.}}
A churn in the indistinguishability sets occurs when users do not send in an \textit{active slot} in their schedules. In our protocol, we assume that the churn occurs only due to unintentional reasons, e.g., the users fail to send due to a network connection problem. Hence, we do not take into account when users fail to adhere to the schedule due to active attacks, such as delaying or dropping messages by an adversary (cf.\ Section~\ref{sec:tm}).\\

\paragraph*{\textbf{Elimination.}}
When a user $u_i$ does not follow the schedule in one \textit{active slot}, the protocol supports two settings:
\begin{itemize}
\item \textit{No chances:} $u_i$ will be removed from her set and no longer be able to publish under her pseudonym. The user is eliminated because she has deviated from the behavior that is obliged by the schedule, i.e., she behaved differently compared to the other users in her set. As a result, the protocol can no longer guarantee indistinguishability for her. If $u_i$ wants to publish posts on the system again, she must join the system as a new user with a new identity. That means she will be part of a new batch and go through the mitigation protocol phases again.

\item \textit{Chances:} It may not be practical to eliminate $u_i$ if she does not follow the schedule in a single \textit{active slot}. Thus, in order to maintain a level of practicality without breaking the indistinguishability, the protocol imposes a delay time $d$ allowing to wait for $u_i$ to send her messages. Therefore, when $u_i$ does not send messages in an \textit{active slot}, the messages sent by other users who belong to the same set of $u_i$ will be delayed. It will be published on the system either when $u_i$ sends her messages before the $d$ period ends or when the waiting time exceeds $d$. 
In the latter case, $u_i$ will be eliminated. In the beginning of the \emph{Communication Phase}, the protocol can assign each user a number of failure times, i.e., a user is allowed to fail to follow the schedule up to this number. \\
\end{itemize}

\paragraph*{\textbf{Updating the schedule.}}
In the previous phase, a schedule $H_x$ is created based on the publishing behavior on pseudonyms during $T_{learning}$. In the \emph{Communication Phase}, the protocol can update $H_x$ to adjust it to the recent history of the publishing behavior. That means $H_x$ in $T_{e+1}$ will be based on the publishing behavior during~$T_{e}$. To accomplish this, the steps below are carried out for each time interval~$T_{e}$:

\begin{itemize}
\item Compute a new instance of $B_{p_j}$ for each pseudonym $p_j$ during the time interval $T_e$ using the equation~\ref{eq:1}.
\item Create $H_x^{Temp}$ using the equation~\ref{eq:5}. 
\item Update $H_x$ to equal $H_x^{Temp}$, if $H_x^{Temp}$ contains at least a certain number of \textit{active slots}.
\end{itemize}

%% file: sections/eval.tex
\section{Evaluation}
\label{sec:eval}
In this section, we analyze realistic user publishing behavior in microblogging settings and assess the efficiency of our proposed mitigation protocol in light of this behavior. A prototype of our protocol is implemented in Python. The user publishing behavior in the prototype is derived from two real-world datasets collected from two popular microblogging platforms, Twitter and Reddit. The batch threshold is set to 5000. The number of slots $w$ in a time interval $T_e$ is 24, and the size of each slot $t_l \in T_e$ is~1 hour (in accordance with related work \cite{hayes16, gaballah2022effectiveness}). Therefore, the \emph{Learning Phase} in our experiments lasts for 24 hours. The message size is assumed to be 1 KB as the messages in microblogging scenarios are typically small, e.g., the text content of a Tweet can contain up to 280 characters or Unicode glyphs \cite{tweetsize}. The number of messages $m$ that a user can send in a time slot is set to 1 (as in \cite{hayes16, riposte, RefRiffle, sarah}).  The number of the clusters/sets $k$ is 15 which is chosen using the elbow method. The minimum size $z$ of every set $S_x$ is set to 50. 

\subsection{Datasets}
We used two datasets in our evaluation. The first dataset is an already existing collection of records extracted from Twitter over the course of the entire month of November 2012 \cite{datasetsrc1}\cite{datasetsrc2}. This dataset contains 22534846 tweets, 6914561 users, and 3379976 topics, referred to as hashtags. The second dataset is collected by us from Reddit for the whole month of October 2021. This dataset contains posts and comments from 1638157 different users and 3403 different topics, referred to as subreddits. Both datasets include a timestamp, a user id, and a topic (hashtag/subreddit) at each record.

\begin{figure*}[t!]
    \centering
    \begin{subfigure}[t]{0.32\textwidth}
        \centering
        \includegraphics[height=44mm]{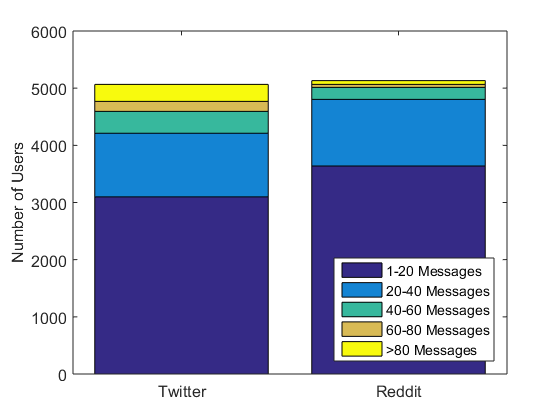}
        \caption{The distribution of users based on total publishing rates.}\label{fig:eval1}
    \end{subfigure}    
    \begin{subfigure}[t]{0.32\textwidth}
        \centering
        \includegraphics[height=44mm]{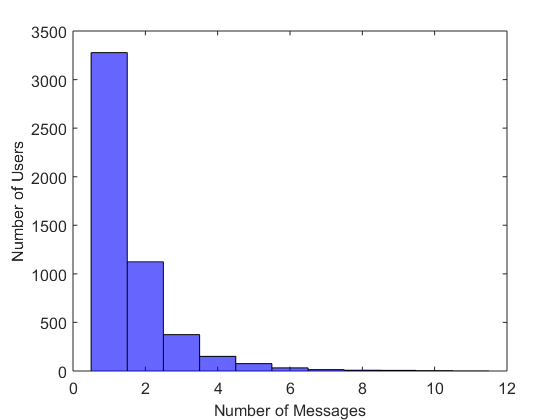}
        \caption{Twitter: Histogram of the publishing rates during the \emph{Learning Phase}.}\label{fig:eval2}
    \end{subfigure}
    \begin{subfigure}[t]{0.32\textwidth}
        \includegraphics[height=44mm]{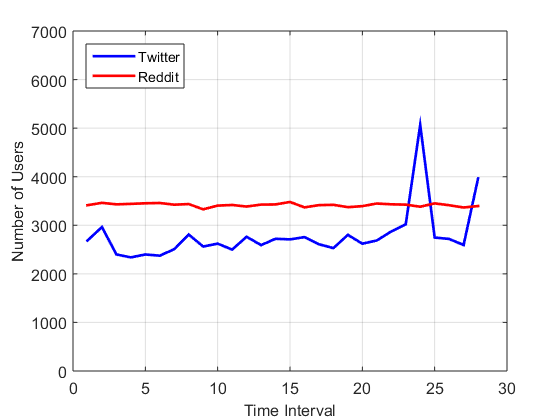}
        \caption{The number of users who do not publish during each Time Interval.}\label{fig:evalnotsending}
    \end{subfigure}

    \begin{subfigure}[t]{0.32\textwidth}
        \includegraphics[height=44mm]{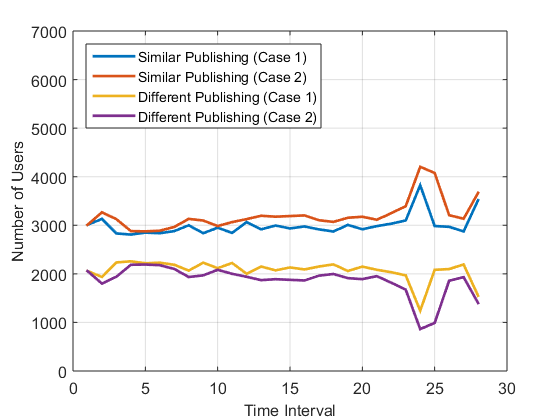}
        \caption{Twitter: The change in user publishing behavior over intervals.}\label{fig:eval3}
    \end{subfigure}
           \begin{subfigure}[t]{0.32\textwidth}
        \includegraphics[height=44mm]{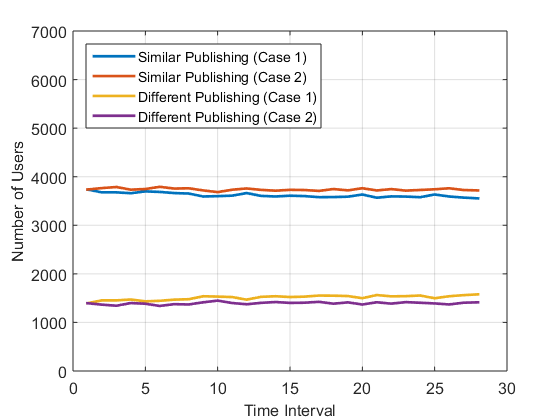}
        \caption{Reddit: The change in user publishing behavior over intervals.}\label{fig:eval4}
    \end{subfigure}
       
    \caption{Analysis of user publishing behavior}
    \label{fig:part1}
\end{figure*}

\subsection{Analysis of Realistic User Publishing Behavior}
\label{subsec:anal}
User publishing behavior has a considerable impact on the various phases of the protocol. Thus, we examined user behavior in batches from the two datasets. Figure~\ref{fig:eval1} displays the total publishing rates of users over the course of a month, where the batch threshold is 5000. As shown in the figure, the majority of Twitter and Reddit users sent between 1 and 20 messages during the month. Twitter users publish at a higher rate than Reddit users; for example, the number of users who send more than 60 messages on Twitter is much higher than on Reddit. The difference in publishing rates between Twitter and Reddit users is to be expected, given the two platforms' different service models. The distribution shown in Figure~\ref{fig:eval1} was found to be nearly the same for various batch threshold values (we ran the analysis for values of 1000, 2000, 3000, 5000, and 7000). 

Figure~\ref{fig:eval2} shows the low publishing rates of Twitter users during the \emph{Learning Phase}. That means the vectors created during this phase typically have a limited number of slots observed with real messages. The low publishing rates during the \emph{Learning Phase} were also noticed among Reddit users. Additionally, Figure~\ref{fig:evalnotsending} illustrates that a large number of users do not publish during each time interval. That, again, emphasizes the low publishing rates of users.

We also looked at how the user publishing behavior evolves over time. This investigation is necessary to understand whether a single fixed schedule for each set is sufficient or whether the schedule should be updated regularly. To accomplish this, we considered two cases:

\begin{itemize}
\item Case 1: The user's publishing behavior during the first interval $T_1$ (which represents the \emph{Learning Phase} in our protocol) is compared to her behavior during the subsequent intervals $T_2,T_3,...T_v$. If they are similar, then a schedule created based on her behavior in the \emph{Learning Phase} will be appropriate for the communication during $T_e$. However, if they are different, the schedule will be less suitable for communication during $T_e$.

\item Case 2: The user's publishing behavior during every two consecutive intervals $T_e$ and $T_{e+1}$ is compared. If they are similar, a schedule created based on her behavior during interval $T_{e}$ is suitable for communication during the interval $T_{e+1}$.

\end{itemize}

The simple matching coefficient (SMC) is used in both cases to determine the similarity and dissimilarity of the publishing behavior. Figures~\ref{fig:eval3} and~\ref{fig:eval4} show the number of users who exhibit similar or different behavior in each time interval based on the aforementioned cases; the results are illustrated for Twitter and Reddit batches, respectively. For instance, in Figure~\ref{fig:eval3}, during time interval number 5, there are roughly 2900 users with publishing behavior similar to their publishing behavior in the first interval (the \emph{Learning Phase}) based on Case~1. Two publishing behavior vectors were deemed similar in our analysis if they were identical or only differed in one slot.

The two figures depict that, in both Case 1 and Case~2, there are more users with similar publishing behavior than those with different behavior. Hence, it is a worthwhile endeavor to create schedules based on the users’ communication history. Case~2 demonstrates results for similarity higher than Case 1. Although the gap between the results of the two cases is not significant, it still does imply that updating the schedules for intervals might result in more representative schedules and, thus, less bandwidth and latency overhead.

We think that the reason for the dynamic nature of the results in Figure~\ref{fig:eval3} versus those in Figure~\ref{fig:eval4} is that the publishing behavior of Twitter users is more triggered by hot topics and trends, i.e., users tend to publish more or less depending on the presence of hot topics. That does not appear to be the case on Reddit, where users seem to be more consistent in their publishing behavior.

\begin{figure*}[t!]
    \centering
     \begin{subfigure}[t]{0.32\textwidth}
        \includegraphics[height=44mm]{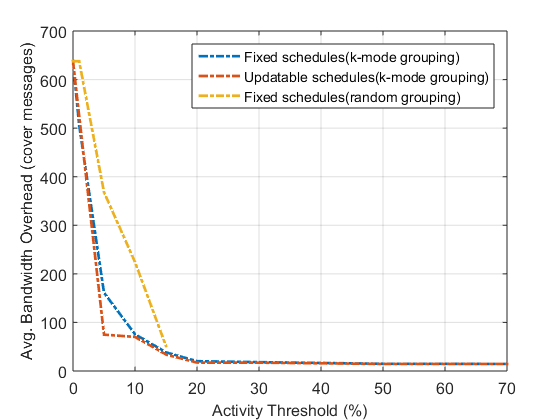}
        \caption{Twitter: The avg. bandwidth overhead per user.}\label{fig:eval5}
    \end{subfigure}
      \begin{subfigure}[t]{0.32\textwidth}
    \includegraphics[height=44mm]{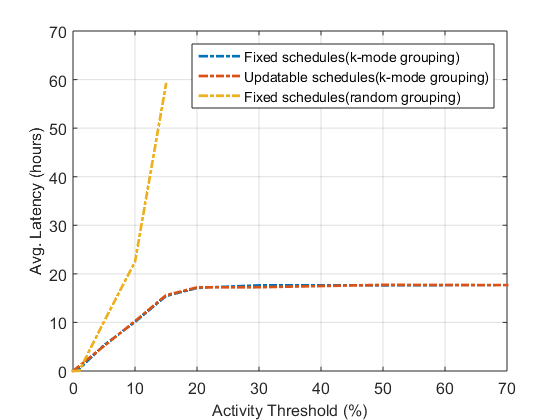}
        \caption{Twitter: The avg. message latency.}\label{fig:eval7}
    \end{subfigure}
    
    \begin{subfigure}[t]{0.32\textwidth}
        \includegraphics[height=44mm]{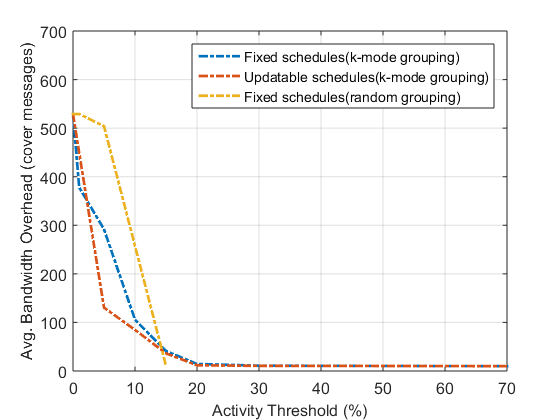}
        \caption{Reddit: The avg. bandwidth overhead per user.}\label{fig:eval6}
    \end{subfigure}
      \begin{subfigure}[t]{0.32\textwidth}
    \includegraphics[height=44mm]{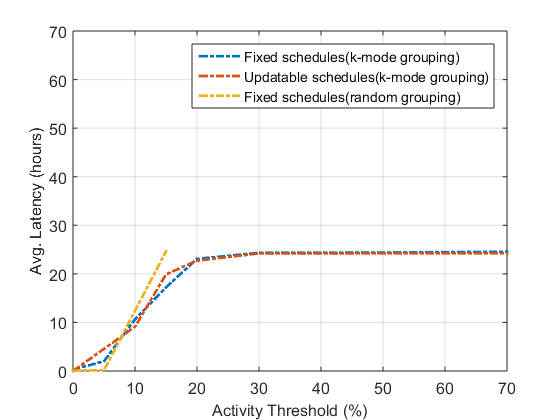}
        \caption{Reddit: The avg. message latency.}\label{fig:eval8}
    \end{subfigure}
    \caption{The impact of activity threshold
on bandwidth overhead and latency}
    \label{fig:part2}
\end{figure*}

\begin{figure*}[t!]
    \centering
         \begin{subfigure}[t]{0.32\textwidth}
        \includegraphics[height=44mm]{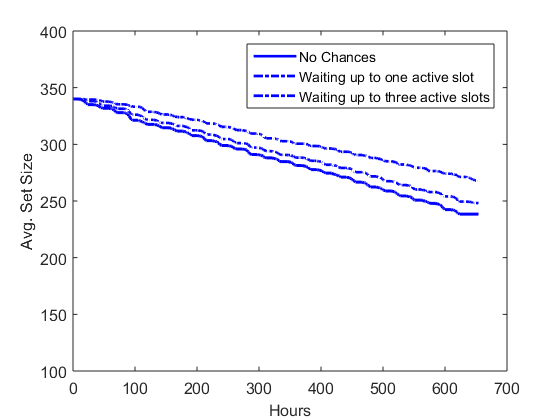}
        \caption{Twitter: Churn rate is 30\%.}\label{fig:eval9}
    \end{subfigure}
        \begin{subfigure}[t]{0.32\textwidth}
        \includegraphics[height=44mm]{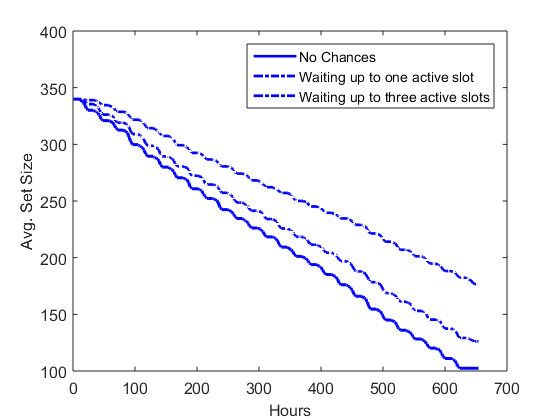}
        \caption{Twitter: Churn rate is 70\%.}\label{fig:eval11}
    \end{subfigure}
    
    \begin{subfigure}[t]{0.32\textwidth}
        \includegraphics[height=44mm]{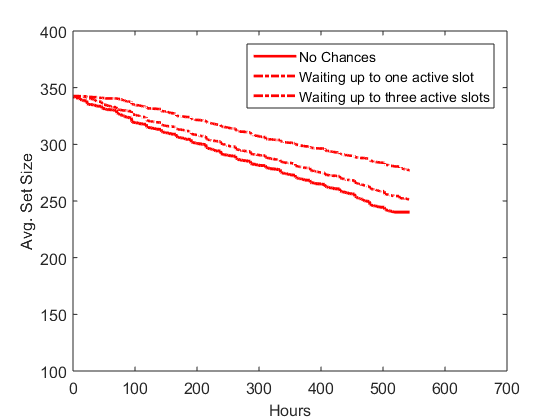}
        \caption{Reddit: Churn rate is 30\%.}\label{fig:eval10}
    \end{subfigure}
           \begin{subfigure}[t]{0.32\textwidth}
        \includegraphics[height=44mm]{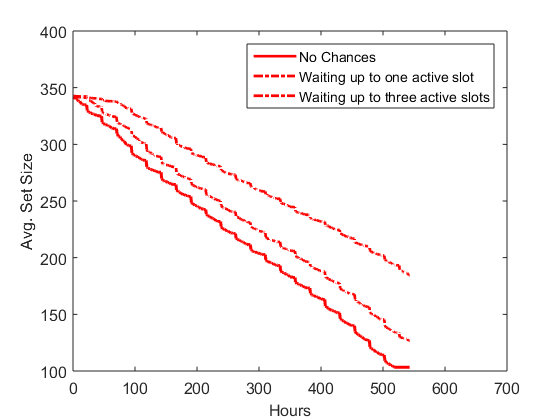}
        \caption{Reddit: Churn rate is 70\%.}\label{fig:eval12}
    \end{subfigure}

    \caption{The impact of churn on the average indistinguishability set size.}
    \label{fig:part3}
\end{figure*}


\subsection{Bandwidth and Latency Overhead} 
Our protocol aims at protecting users against intersection attacks by requiring users from the same set to communicate indistinguishably by following a schedule. However, communication based on schedules can be costly in terms of bandwidth and latency overhead. Since the activity threshold $q$, defined in Section~\ref{sec:p4}, affects the amount of the overhead, we tested several values for this parameter to assess the efficiency of the schedules. In Figure~\ref{fig:part2}, for instance, when the threshold value is 10\%, it means that a time slot $t_l$ is an \textit{active slot} (i.e., $H_{x,t_l}=1$) if at least there are 10\% of the publishing vectors having a value of 1 at $t_l$. 

The bandwidth overhead and latency that the schedules impose are influenced by how the users are grouped. Therefore, we compared how effective the schedules are when the sets are formed using the k-mode algorithm and when they are created randomly (i.e., users are randomly divided into sets during the grouping phase). The random grouping was carried out ten times, and the bandwidth and latency overhead results of each value of $q$ were then averaged. We found that the value of $q$ must be 15\% or lower in order to generate schedules for random groups. Even when it is 15\%, for some groups (usually 5 to 8 out of the 15 groups), the schedules cannot be created. The behavior vectors in each group are so different from one another, so it is challenging to discover overlapping \textit{active slots} between the vectors. That makes it impossible for the protocol to produce schedules when the value of $q$ is greater than~15\%. Nevertheless, that is not the case when the behavior vectors are grouped into sets based on similarity using the k-mode algorithm.

Additionally, we evaluated the overhead when the schedules are fixed and when they are updated during the \emph{Communication Phase}. In our experiments, the schedule of a set $S_x$ is only updated when the new schedule $H_x^{Temp}$ contains at least two \textit{active slots}. We discovered that updatable schedules are feasible only when the k-mode, not random grouping, is employed for grouping. The substantial disparities between the behavior vectors in each group are to blame once more for this.

\paragraph*{\textbf{Bandwidth overhead.}} The bandwidth overhead per user is calculated in our evaluation by counting the total number of cover messages sent by a user during the \emph{Communication Phase}. Figures~\ref{fig:eval5} and~\ref{fig:eval6} depict the average total bandwidth overhead per user. The fixed and updatable schedules based on k-mode grouping have notably lower bandwidth overhead than the schedules based on random grouping. That is to be expected because the k-mode algorithm provides sets with more similar publishing behavior than random grouping. 

In both Figures~\ref{fig:eval5} and~\ref{fig:eval6} , the results show an inverse relationship between the bandwidth overhead and the threshold $q$. Hence, increasing the value of $q$ leads to lower overhead, and vice versa. This aligns with intuition because a low threshold implies that if a set contains a few publishing vectors with a value of 1 at $t_l$, the~protocol will consider this sufficient to make $t_l$ an \textit{active slot}. Accordingly, the resulting schedule pushes all other users in the set (those with vectors with a value of 0 at $t_l$) to send cover messages in order to achieve indistinguishability. 

As illustrated in the figures, the average bandwidth overhead stabilizes at a value of 20\% and higher for fixed and updatable schedules that are based on k-mode. This indicates that there is no discernible change in the schedules at these values. The reason for this is that users' publishing rates are low, both overall and during the \emph{Learning Phase}, indicating sparse slots with a value of 1 in the publishing vectors. That makes it difficult to reach the high threshold value required to consider a slot $t_l$ as an \textit{active slot}. As a result, there is no need to increase the threshold value beyond 20\%, as this does not lead to any further optimization. Another interesting observation is that the updatable schedules can introduce less bandwidth overhead than fixed schedules only if $q$ is less than 15. This observation holds true for both the Twitter and Reddit results. This could also be due to the low publishing rates, which means that no further improvements can be made when increasing the threshold regardless of whether or not the schedules are updated. Another reason is that the number of users in the batches who do not send messages at all in each interval is very large. At each interval, more than 2500 Twitter users and around 3500 Reddit users do not send any messages (cf.\  Figure~\ref{fig:evalnotsending}). As a result, most new schedules typically contain only zero values (i.e., \textit{inactive slots}), and the protocol disregards them because they lack at least two \textit{active slots}. That means changes in the schedules do not often happen, which makes the results of the fixed and updatable schedules not that different.

\paragraph*{\textbf{Latency.}} Figures~\ref{fig:eval7} and~\ref{fig:eval8} demonstrate the average message latency introduced by the schedules during the \emph{Communication Phase}. The message latency is calculated as the time between generating a message on the user's side and publishing it on the system's side. As shown in the figures, the fixed and updatable schedules that are based on k-mode introduce lower latency than those that are based on random grouping. The latency results of fixed schedules are similar to the updatable schedules. The reasons behind this are similar to those explained in the bandwidth section. 

In contrast to the results of the bandwidth overhead evaluation, the latency has a direct relationship with the threshold $q$. That is, increasing the threshold value increases latency, and vice versa. This is understood because when the threshold value is high, a larger number of publishing vectors with a value of 1 at $t_l$ are required to make $t_l$ an \textit{active slot}; this is also difficult given the low publishing rates of users in the two datasets. As a result, the number of \textit{active slots} may be limited, causing many messages to wait on the user's side for some time before being sent. 

When we compare the results between the two datasets, the average message latency is higher on Reddit than on Twitter for larger values of the threshold , as shown in Figures~\ref{fig:eval7} and~\ref{fig:eval8}. That makes sense when looking at the publishing rates since Reddit users have lower publishing rates than Twitter users. Again, from the results of both Twitter and Reddit, we see stability at 20\% and higher, so there may be no need to raise the threshold value after~20\%.

The bandwidth overhead and latency have an inverse relationship, which means that low latency indicates fast message publication at the expense of high bandwidth overhead, and vice versa. Therefore, the threshold value should be chosen in such a way that users in an indistinguishability set get a good trade-off between bandwidth overhead and latency. A low threshold should be chosen if low latency is critical, whereas a high threshold can be used when low bandwidth overhead is most important. \footnote{The value of $q$ can be determined separately for each set to serve the needs and preferences of the users in that set.} 


\subsection{Anonymity under Network Churn}  
Naturally, the network will experience churn as some users might be unable to send during \textit{active slots}. Accordingly, the size of the indistinguishability set will inevitably decrease over time. However, the set size should not degrade quickly. We consider the average indistinguishability set size during the communication phase to assess the impact of churn on anonymity. We simulate churn per set by randomly selecting users from the set to not adhere to their schedule during randomly selected time slots. The number of these selected users from each set is referred to as the churn rate. When the rate is, say 50\%, it means that 50\% of the users in the set will be chosen at random to ignore the schedule during randomly selected time slots. Figures~\ref{fig:eval9} to~\ref{fig:eval12} depict the average indistinguishability set size when the churn rate per indistinguishability set is 30\% and 70\%. As expected, the average set size decreases faster as the churn rate per set increases.

We compared the results when the protocol does not give users chances to the results when chances are given. We considered two cases for the chances. The first case is when a user fails to send during an \textit{active slot}, the protocol waits until the next \textit{active slot}. If the user does not send the messages from the previous and new slots during this slot, the user will be removed from the set. In the second case, we increase the waiting time to force the protocol to wait for three consecutive \textit{active slots}; if the user has not sent the required messages by then, she is eliminated. Giving users chances when they miss sending in an \textit{active slot}, as shown in the figures, significantly slows down the degradation in the set size, especially when the churn rate is high or the waiting period is increased. In Figure~\ref{fig:eval11}, for example, when the churn rate is 70\% and the waiting time is up to three \textit{active slots}, the average set size drops to around 180 by the end of the simulated \emph{Communication Phase}, which is greater than the set size when no chances are given. Even if no such chances are provided and the churn rate is high, users will be protected by sufficiently large indistinguishability sets. Obviously, if the protocol waits longer for users who miss their schedules to send the messages, the publishing of the users' messages may be delayed, i.e., the latency may increase.

We discovered that the majority of users are typically inactive (i.e., not sending real messages) between 12 a.m. and 6 a.m. Thus, the time slots located during these times are not active in most schedules. The reason for this could be that users are not active during these hours due to sleeping. Since users can only be eliminated from their set during \textit{active slots}, any decrease in the set size occurs only during these slots. Therefore, the size of the set remains constant during \textit{inactive slots}.

%% file: sections/diss.tex
\section{Discussion}
\label{sec:diss}
In this section, we discuss some additional protocol settings.

\paragraph*{\textbf{Latency during the arrival phase}}
A batch of a particular size is required in our protocol to begin the \emph{Learning Phase}. The joining rate of new users determines the latency for gathering the batch. If the joining rate is high enough, the batch threshold will be met rapidly, resulting in low latency. However, if users join slowly, the latency until the \emph{Learning Phase} starts will be high. Despite this limitation, we believe that waiting for a specific number of new users is better than waiting for a certain period of time to collect new users. The former approach is better because it guarantees a large batch of new users from the start, which aids in creating large anonymity sets for users when we group them based on publishing behavior. Therefore, to avoid waiting for a long time, or even forever, the protocol can wait until the number of new users reaches the batch threshold or wait up to a specific amount of time.

\paragraph*{\textbf{Invalid sets.}}
To ensure a minimum level of indistinguishability, the protocol must ensure that the size of each indistinguishability set $|S'|$ is at least $z$. However, during the grouping phase, some of the clusters produced by the $k$-mode algorithm may be smaller than the threshold. These clusters are identified as \textit{invalid sets}. Following the \emph{Grouping Phase}, pseudonyms assigned to \textit{invalid sets} will be kept until the protocol receives a new batch. The protocol groups these delayed pseudonyms with the new batch's pseudonyms. When a \textit{valid set} (i.e., a set with a size larger than or equal to $z$) contains pseudonyms from different batches, the users of the pseudonyms in this set should publish under new pseudonyms during the \emph{Communication Phase}. That is crucial to prevent $\mathcal{A}$ from partitioning users belonging to the same set into subsets or even de-anonymizing some of them using a timing attack based on when the pseudonyms began to receive messages.

\paragraph*{\textbf{Leaving the system.}}
Most microblogging systems allow users to delete their accounts/pseudonyms when they no longer want to use the system. However, stopping the use of the system causes churn in the indistinguishability sets. Therefore, to ensure that users are indistinguishable after leaving the system, users must first inform the system that they wish to delete their pseudonyms. The users need to stick to the schedule until the protocol notifies them that their pseudonyms have been deleted. Once they receive this notification, they can stop using the system . The protocol deletes the user's pseudonym when there are at least a certain number of other users in the set who also want to delete their pseudonyms.

%% file: sections/relatedwork.tex
\section{Related Work}
\label{sec:rw}
\paragraph*{\textbf{Anonymous microblogging systems.}}
Several anonymity systems have been proposed throughout the last years to support the microblogging scenario. In these systems, sender anonymity is achieved using various techniques such as \textit{mixnets} (Atom \cite{atom}), \textit{DCnets} (Dissent \cite{RefDissent}), \textit{private information retrieval} (Blinder\cite{RefBlinder}, Riposte \cite{riposte}, 2PPS \cite{sarah}, Spectrum \cite{spectrum}), and \textit{random forwarding} (AnonPubSub \cite{anonpubsub}). For receiver anonymity, most of the proposed systems depend on the concept of broadcasting messages to all users \cite{RefBlinder, RefDissent, riposte, atom, spectrum}, which results in high network overhead. Since broadcasting is not suitable for users with limited bandwidth, systems like \cite{anonpubsub, giakkoupis, RefRiffle, sarah} have addressed this issue by enabling anonymous multicast communication. 

\paragraph*{\textbf{Traffic-analysis attacks.}}
There are several types of traffic analysis attacks, such as timing attacks, intersection attacks, and statistical disclosure attacks. Tor, which is by far the most widely used anonymity system, is vulnerable to the three previously mentioned attacks \cite{RWTor2,  statistical, openlimits}. Signal, a popular privacy-preserving instant messaging application, is also susceptible to a statistical disclosure attack that can effectively deduce the relationship between the sender and the recipient of an end-to-end encrypted message stream in the application \cite{signalattack}. Although mixnets are typically known for their ability to withstand traffic analysis attacks, there are studies \cite{Troncoso_perfectmatching,openlimits,2ssda} that evinced the effectiveness of statistical disclosure attacks against mix-based systems, especially when they support full bidirectional communications \cite{2ssda}. On anonymous email networks, statistical disclosure attacks based on the Expectation-Maximization algorithm were successful as well \cite{portela2}. In \cite{gaballah2022effectiveness}, intersection attacks were demonstrated to be extremely effective on anonymous microblogging.

\paragraph*{\textbf{Mitigation techniques.}}
Sending cover traffic throughout the whole mix network has been shown to be able to prevent some traffic analysis attacks. Nonetheless, it cannot overcome powerful attacks like statistical disclosure attacks, and intersection attacks \cite{berthold03,gaballah2022effectiveness}. Many papers like \cite{twocent,cmix,sarah,RefRiffle,RefDissent} proposed protecting
against intersection attacks by requiring all users who participate in the systems to have similar communication behavior, i.e., all users join the system at the same time, send at the same time slots, and have the same sending rates. Nevertheless, this requirement is not realistic. A method was proposed in \cite{hayes16} for forming possibilistic anonymity sets by grouping users based on their communication behavior. However, possibilistic anonymity sets do not ensure strong anonymity (i.e., indistinguishability) as it only ensures plausible deniability. Another solution was proposed in \cite{wolinsky13} to create anonymity sets that ensure indistinguishability. Nevertheless, this solution groups the users randomly into sets; hence, it is not efficient.

%% file: sections/conclusion.tex
\section{Conclusion \& Future Work}
\label{sec:confw}
In this paper, we propose a protocol for mitigating intersection attacks in anonymous microblogging systems. Our protocol addresses the unrealistic requirement in the literature that all users must commit to send messages all the time to prevent user de-anonymization via intersection attacks. It groups users based on their publishing behavior into sets. Then, for each set, it generates a communication schedule and requires users in the set to adhere to the schedule in order for them to appear indistinguishable from the point of view of an adversary. In our evaluation, we used real-world datasets from Twitter and Reddit to derive realistic user publishing behavior. We examined the users' behavior in these datasets and discovered that the majority of users in both datasets have low publishing rates. Our findings also show that scheduling can significantly reduce bandwidth overhead on the user's side. However, as expected, we have found that the reduction in the bandwidth overhead usually comes at the expense of latency. Therefore, the schedule for each set should be designed in such a way that it optimizes the trade-off between bandwidth overhead and latency based on the needs of the users in that set.  

Future work should focus on testing our protocol on more datasets collected over a longer period. In addition, different ways of creating schedules can be applied and assessed. In particular, more sophisticated methods (e.g., machine learning algorithms) can be employed to design schedules based on predictions of user behavior. As a result, schedules might even be enhanced in terms of bandwidth and latency overhead. Furthermore, our mitigation protocol can be applied in other scenarios, such as anonymous messaging.

\section{Acknowledgements.} This work was partially supported by funding from the German Research Foundation (DFG), research grant 317688284. We would like to thank Tim Grube for his insightful feedback on an earlier version of this work. 